\documentclass[aip,rsi,reprint]{revtex4-1}

\usepackage{graphicx}
\usepackage{color}

\begin{document}


\title{Tiny adiabatic-demagnetization refrigerator for a commercial superconducting quantum interference device magnetometer}


\author{Taku J Sato}
\email[]{taku@tagen.tohoku.ac.jp}
\affiliation{Institute of Multidisciplinary Research for Advanced Materials, Tohoku University, 2-1-1 Katahira, Sendai 980-8577, Japan}

\author{Daisuke Okuyama}
\affiliation{Institute of Multidisciplinary Research for Advanced Materials, Tohoku University, 2-1-1 Katahira, Sendai 980-8577, Japan}

\author{Hideo Kimura}
\affiliation{National Institute for Materials Science, 1-2-1 Sengen, Tsukuba 305-0047, Japan}

\date{\today}

\begin{abstract}
A tiny adiabatic-demagnetization refrigerator (T-ADR) has been developed for a commercial superconducting quantum interference device magnetometer [Magnetic Property Measurement System (MPMS) from Quantum Design].
The whole T-ADR system is fit in a cylindrical space of the diameter $8.5$~mm and the length $250$~mm, and can be inserted into the narrow sample tube of MPMS.
A sorption pump is self-contained in T-ADR, and hence no complex gas handling system is necessary.
With the single crystalline Gd$_3$Ga$_5$O$_{12}$ garnet ($\sim 2$~grams) used as a magnetic refrigerant, the routinely achievable lowest temperature is $\sim 0.56$~K.
The lower detection limit for a magnetization anomaly is $\sim 1 \times 10^{-7}$~emu, estimated from fluctuation of the measured magnetization.
The background level is $\sim 5 \times 10^{-5}$~emu below 2~K at $H = 100$~Oe, which is largely attributable to a contaminating paramagnetic signal from the magnetic refrigerant.
\end{abstract}


\maketitle


\section{Introduction}
Magnetization measurement at low temperatures is one of the most fundamental techniques for condensed matter sciences.
Nowadays, superconducting quantum interference device (SQUID) magnetometers become commercially available, and are widely used for magnetization measurements.
A representative example of such commercial SQUID magnetometers may be the Magnetic Properties Measurement System (MPMS) from Quantum Design~\cite{QD}.
The basic setup of this instrument enables magnetization measurement down to 1.8~K under the magnetic field up to 5~T with the resolution being $\sim 1 \times 10^{-7}$~emu using the standard DC extraction method.
Versatility of MPMS has stimulated a number of developments extending its capability for other types of measurements such as heat capacity~\cite{Kharkovski07}, magnetoelectric properties~\cite{Borisov07}, and even nuclear quadrupole double resonance~\cite{Shroyer11}.

Extending temperature range for the magnetization measurement to much lower side is strongly desired, since many interesting quantum phenomena, such as unconventional superconductivity~\cite{Norman14}, novel heavy fermion magnetism~\cite{Flouquet11}, and quantum spin liquid state~\cite{Balents10}, have been either observed or sought after below 1.8~K.
Exemplified by the cerebrated unconventional superconductor Sr$_2$RuO$_4$, non-Bardeen-Cooper-Schrieffer (BCS) type superconductivity was reported with the transition temperature $T_{\rm c} \simeq 0.9$~K, and has been under intensive investigation to elucidate its superconducting mechanism~\cite{Mackenzie03}.
Another topic may be geometrically frustrated quantum spin systems, where an intriguing quantum disordered state, known as the spin liquid state, has been sought after at low-temperatures, ideally at 0~K~\cite{Balents10}.
Undoubtedly, the magnetization measurement is the most fundamental tool to elucidate magnetic and superconducting properties of materials under investigation, and thus for the cutting-edge condensed-matter science, the ultra-low-temperature magnetization measurement is of crucial importance.

In such an ultra-low-temperature range, several custom-made magnetometers were developed, based on either the $^3$He pumping or $^3$He-$^4$He dilution refrigeration technique~\cite{Paulsen90,Schlager93,Sakakibara94,Dumont02,Nishioka02,Legl10,Martinez-Perez10}.
Magnetization of the sample is measured using inductive methods~\cite{Paulsen90,Schlager93,Dumont02,Nishioka02,Legl10,Martinez-Perez10}, or the Faraday force magnetometry~\cite{Sakakibara94}.
Recent development along this line enables ultra-low-temperature ac susceptibility measurements even on micron-sized magnets by setting the sample as well as SQUID microsusceptometer in the mixing chamber~\cite{Martinez-Perez10}.
Although those elaborated equipments provide high-quality magnetization data, a sample is usually placed in the mixing chamber or the $^3$He bath of the refrigerator\cite{Nishioka02,Martinez-Perez10}, or attached to the mixing chamber~\cite{Paulsen90,Schlager93,Sakakibara94,Dumont02,Legl10}, which is deep inside the cryostat, thermally insulated by several vacuum layers.
Hence, the whole magnetization-measurement system becomes quite complicated with its sample change being a cumbersome task, requiring special expertise in low-temperature physics.
Consequently, such ultra-low-temperature magnetization measurement is possible at only selected laboratories.
Quantum Design also provides a $^3$He insert (iQuantum) for the MPMS magnetometer as an option for the ultra-low-temperature magnetization measurements, which extends the measurement range down to $T = 0.42$~K~\cite{QD}.
There has been continuous effort to increase sensitivity of the magnetization measurement using such a $^3$He insert~\cite{Sato13}.
However, as it also uses $^3$He-pumping technique, the precious and expensive $^3$He gas, as well as the complex gas handling system, is required, making this option to be rather unaccessible for general users.
Therefore, it has been highly desired to develop a much simplar apparatus for magnetization measurements at ultra-low temperatures based on a fundamentally different refrigeration technique.

In this work, we utilize the adiabatic-demagnetization refrigeration (ADR) technique to realize an ultra-low-temperature refrigerator to be used in combination with a commercial MPMS magnetometer.
ADR is based on the magnetocaloric effect of paramagnetic materials, called magnetic refrigerants.
In classical thermodynamics, the adiabatic temperature change due to the magnetocaloric effect is given as~\cite{Wikus14}:
\begin{equation}\label{eq:magnetocaloric}
\frac{{\rm d}T}{{\rm d}H} = - \frac{T}{C_H}\left ( \frac{\partial M}{\partial T} \right )_H,
\end{equation}
where $T$ and $H$ are temperature and magnetic field, whereas $M$ and $C_H$ are magnetization and specific heat of the refrigerant.
The temperature change is related to a change of the isothermal magnetic-entropy $S$ through the Maxwell relation:
\begin{equation}\label{eq:Maxwell}
\left ( \frac{\partial S}{\partial H} \right )_T = \left ( \frac{\partial M}{\partial T} \right )_H,
\end{equation}
and hence a refrigerant with high magnetic-entropy change is used for efficient ADR.
ADR is robust solid-state technique that works without gravity, in contrast to the $^3$He-$^4$He dilution refrigeration, and hence is widely used in astronomy applications.
Indeed, a number of reports may be found in literature, such as ADR for a high-resolution X-ray microcalorimeter~\cite{Shirron16,Hoshino12}, and infrared bobometer~\cite{Timbie90}, to note a few.
Utilization of ADR for small laboratory experiments has also been discussed~\cite{Hagmann95}.

To realize an ultra-low-temperature ADR for magnetization measurements using MPMS, there are, however, specific difficulties.
The most serious issue is the space restriction; MPMS sample space is only 9~mm in diameter, and thus all the necessary parts, including a vacuum chamber for thermal insulation, have to be inserted into this restricted sample space.
This results in stringent limitation for magnetic refrigerant volume.
Secondly, a magnetic refrigerant has to be placed sufficiently away from the pickup coil of MPMS when measuring magnetization, to reduce unwanted background signals from a paramagnetic response of the refrigerant.
Thirdly, the whole system (except for the magnetic refrigerant) has to be made of non-magnetic materials in order to increase magnetization-measurement sensitivity.
In this article, we describe the newly developed tiny adiabatic-demagnetization refrigerator (T-ADR), and the results of magnetization measurements at ultra-low temperatures carried out in combination with MPMS.
All the necessary functions, such as the magnetic refrigerant, sorption pump, temperature sensor, and sample, are situated in a small cylindrical space of the diameter $8.5$~mm and the length $250$~mm.
T-ADR does not use the $^3$He gas; this makes T-ADR not only cost-effective, but more importantly very simple and easy-to-use, as no outside gas handling system is necessary.
The routinely achievable lowest temperature is $\sim 0.56$~K.
The outer wall of the vacuum chamber is made of non-magnetic and non-metallic polypropylene, which results in significant improvement of the sensitivity.
It was found that the random fluctuation in measured magnetization is approximately $\sim 1\times 10^{-7}$~emu, which is almost the same as that of the DC extraction measurement with the standard ($T > 1.8$~K) setup of MPMS.
The background level below 2~K is relatively high as $\sim 5\times 10^{-5}$~emu.
This is mainly due to a contaminating paramagnetic signal from the magnetic refrigerant, and can be removed using separately measured background data.

\section{Instrument Design and operation}

\begin{figure}
\includegraphics[scale=0.32, angle=-90, trim={2cm 2cm 5cm 0cm}]{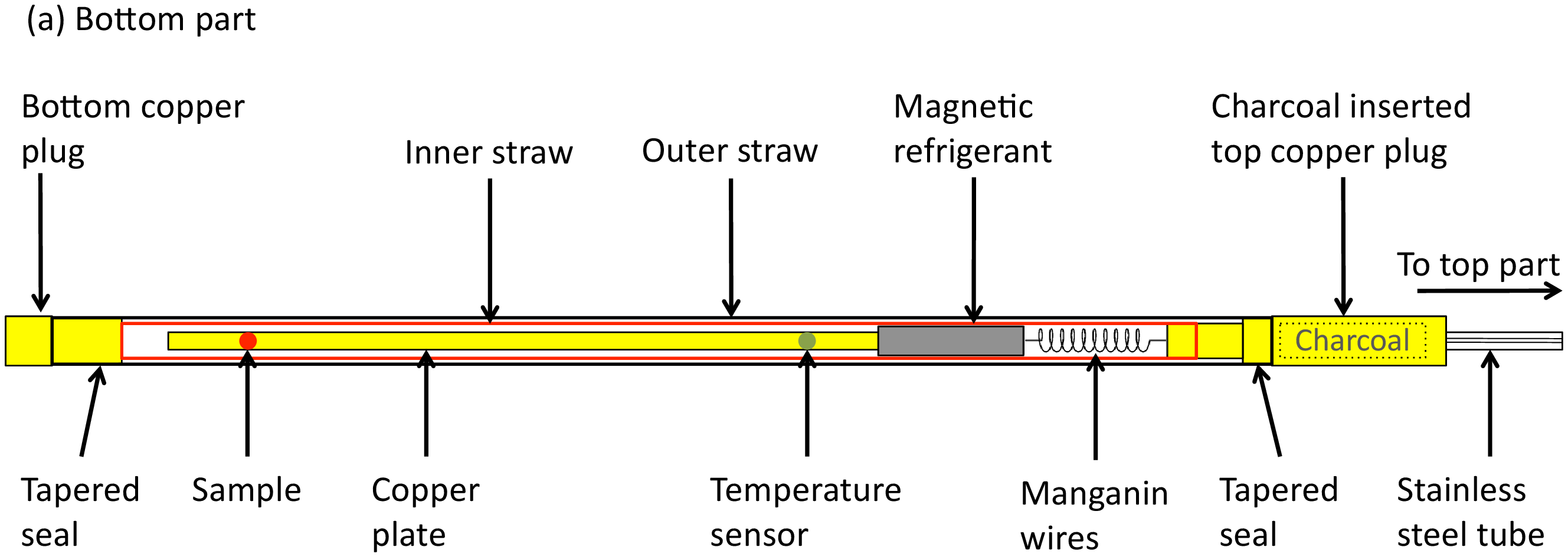}\\
\includegraphics[scale=0.32, angle=-90, trim={2cm 2cm 5cm 0cm}]{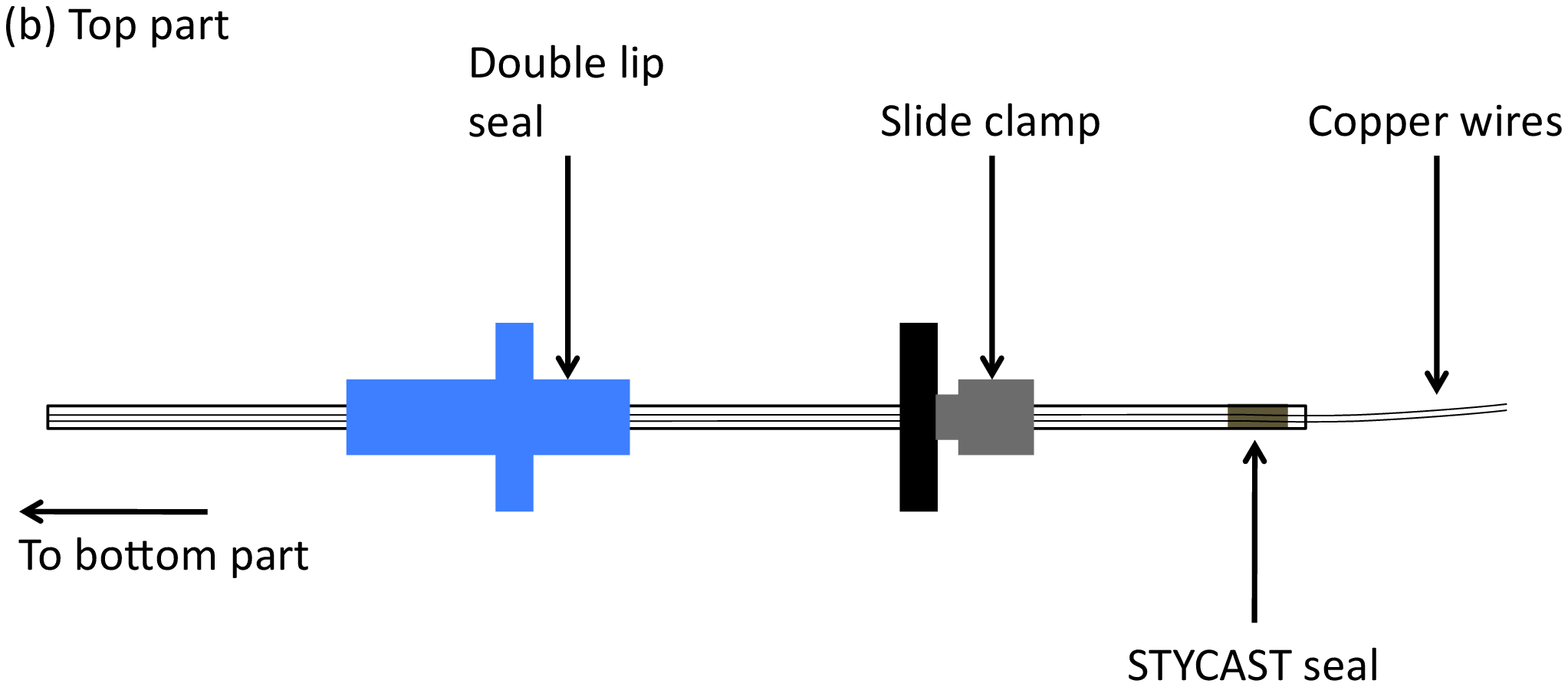}\\
\caption{(Color online) Schematic illustrations of (a) presently developed T-ADR, which is attached to the bottom of the long stainless-steel hanging rod, and (b) top part of the stainless steel rod.
}
\end{figure}

\begin{figure}
\includegraphics[scale=0.48, angle=0, trim={2cm 9cm 2cm 9cm}]{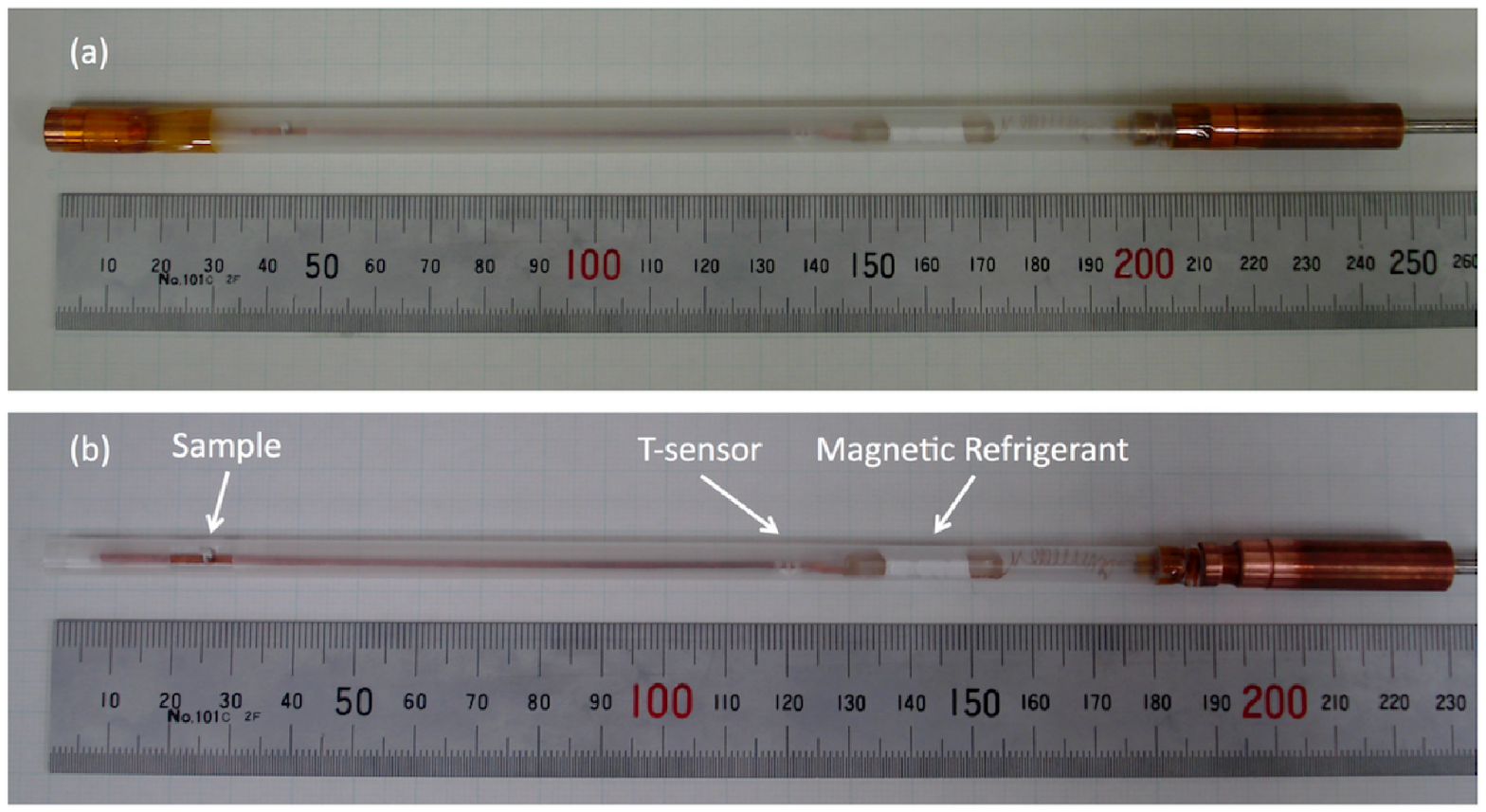}\\
\caption{(Color online) (a) Photograph of developed T-ADR.
Semi-transparent drinking straw is used as the outer vacuum-chamber wall, so that inside can be seen.
(b) Photograph of the inner tube.
The sample, temperature sensor, and magnetic refrigerant are indicated by arrows.
}
\end{figure}

The schematic drawing of T-ADR is shown in Fig.~1, whereas its photographs are presented in Fig.~2.
A long tube with the outer diameter of 3~mm and thickness of 0.3~mm, made of non-magnetic stainless steel, is used as the hanging rod for the T-ADR system.
The top of the tube, shown in Fig.~1(b), is sealed with epoxy (STYCAST 1266), through which electrical leads (four twisted pairs of copper wires with diameter 0.1~mm) are extracted.
At the bottom of the stainless steel tube, shown in Fig.~1(a), a top plug of T-ADR, made of high purity copper (99.9 wt\%), is fixed with silver solder.
Inside the top plug, powder of activated charcoal ($\sim 0.5$~grams) is inserted, which acts as a sorption pump at low temperatures.
The bottom of the top plug is shrunk to 6~mm in diameter, where a drinking straw of the diameter $d \sim 6$~mm (inner drinking straw), made of polypropylene, is attached.
Inside the inner drinking straw, we place a thin copper plate (thickness $t = 0.1$~mm and width $w = 3$~mm), on which the magnetic refrigerant, temperature sensor, and the sample are fixed.
As the magnetic refrigerant, we use the single crystalline Gd$_3$Ga$_5$O$_{12}$ garnet (GGG) ($\sim 2$~grams), whereas for the temperature sensor, we use a commercial ruthenium oxide (RO) chip resister (KOA 1~k$\Omega$), calibrated to the Cernox sensor from Lakeshore.
Among the four twisted pairs of the copper wires inserted in the long $d = 3$~mm tube, two pairs are used for temperature measurement, whereas the others are for auxiliary purposes, occasionally used for electric resistivity measurement.
The copper wires are thermally anchored to the top copper plug, and then extended by the Manganin wires (diameter 0.1~mm) before touching to the low-temperature Cu plate to reduce the heat inflow.
The inner straw is then covered by an outer drinking straw of $d \sim 8$~mm, made of polypropylene, which acts as an outer wall of the vacuum chamber.
The bottom end of the outer straw is sealed by a bottom copper plug.
For vacuum sealing, the contact surfaces of the top and bottom plugs are slightly tapered, and the Apiezon N grease is used as a sealant between the surfaces and straw.

Considerable level of background was anticipated, originating from a large paramagnetic response of the magnetic refrigerant.
The temperature sensor also shows weak but finite magnetization, which smears the signal from a sample.
Hence, the GGG and temperature sensor should be placed as far as possible from the sample position.
By running several background runs with changing the refrigerant position, we found that the background can be reasonably reduced by placing the refrigerant at least 100~mm away from the sample position.
In the present T-ADR setup, we place the refrigerant approximately 110~mm away from the sample position, and the temperature sensor is situated closer to the refrigerant.

The operation of T-ADR is quite simple.
The whole T-ADR rod is inserted into MPMS at room temperature with the magnetic refrigerant placed at 40~mm above the center of the MPMS superconducting magnet.
At this position, the magnetic field is still almost the same as that at the center of the magnet.
Then, the MPMS system temperature is reduced to 20~K, where magnetic field of 5~T is applied to the magnetic refrigerant.
Next, the MPMS system temperature is decreased to 10~K and waited for at least half an hour for the temperature equilibration of the whole MPMS system.
The MPMS system is further cooled down to the base temperature 2~K.
It may be noteworthy that the system should be relatively quickly cooled around 8~K, in order to sufficiently cool the sample in T-ADR before the charcoal sorption pump starts working.
It is usual for the sorption pump to fully working when the MPMS system becomes 4~K, which can be see by the drastic change of cooling rate of the RO sensor temperature.
Once the RO sensor temperature becomes less than 4~K (typically $\sim 2$~K), the magnetic field is turned off.
The RO sensor temperature starts to fall immediately.
Once the RO sensor temperature reaches minimum temperature (less than 0.5~K), the whole T-ADR system is raised by 70~mm, by which the magnetic refrigerant is moved out from the MPMS superconducting magnet, and instead the sample is moved to the center of the magnet.
Then, magnetic field up to 5~kOe is applied to the sample, and magnetization is measured using the usual DC extraction method, as temperature gradually increases at a rate of $\sim 0.008$~K/min as shown later.
(See also Appendix~\ref{appendixTempField} for the effect of magnetic field to the temperature.)

Some notes for material selection may be given here.
First, 99.9 wt\% high purity copper is used for the top and bottom plugs because of less magnetic impurity contained in the material.
It was reported earlier that alloys such as Brass contain unacceptable amount of magnetic impurities~\cite{Sunderland08,Sunderland09}, whereas soft materials such as pure aluminum are difficult to be machined, and tend to fail in vacuum sealing.
Secondly, we use the drinking straw as the material for fixing all the ultra-low-temperature parts.
In the literature, the thermal conductivity of the polypropylene was reported to be $\sim 30~\mu$W/cm K at $T \sim 1$~K~\cite{Barucci02}.
Therefore, for the presently used straw of $d \sim 6$~mm and $t \sim 0.15$~mm, the heat flow from the copper plug to the magnetic refrigerant is approximately $1~\mu$W, which is considerably smaller than the specific heat of the magnetic refrigerant at low temperatures.
It may be noted here that the heat flow through the four twisted pairs of the Manganin wires is estimated as 0.1~$\mu$W for the current setup, and hence is negligible.
We also use the drinking straw for the outer wall of the vacuum chamber.
It is our surprise that the high vacuum necessary to thermally insulate the ultra-low-temperature parts from the 2~K atmosphere, which is typically $10^{-6}$~Torr or less, can be maintained by the thin polypropylene straw; rough-pumped environment in the MPMS sample tube with pressure $\sim 1$~Torr or less may help keeping the high-vacuum inside the thin straw.

Finally, a comment on the selection of the magnetic refrigeration material may be given here.
For the present T-ADR system, there are two contradicting requirements for the magnetic refrigerant.
The magnetic ordering temperature of the refrigerant should be as low as possible (ideally $T = 0$~K) to realize a lower temperature.
On the other hand, we can only use a small magnetic refrigerant for the present T-ADR system due to the tight spatial restriction of the MPMS sample tube ($d = 9$~mm).
Hence, the refrigerant material has to have a large magnetic-entropy change as seen in Eqs.~\ref{eq:magnetocaloric} and \ref{eq:Maxwell}, requiring high magnetic-ion density and large spin quantum number, both of which make the magnetic ordering temperature rather higher.
Compromising the two contradicting requirements, we select GGG as the magnetic refrigerant in the present setup~\cite{Wikus14}.
Availability of single crystals is also an advantage of GGG, ensuring high thermal conductivity necessary to efficiently cool the low-temperature part.
From the earlier work on the thermodynamic properties of GGG~\cite{Numazawa06}, the entropy change $\Delta S$ at $T = 2$~K of 2~grams GGG ($\sim 0.002$~mol) is $\sim 0.085$~J/K on demagnetizing from $H = 5$~T to 0~T, which is apparently sufficient to cool small sample and thin Cu plate.
The lowest achievable temperature is indeed set by the specific heat peak of GGG; due to the development of antiferromagnetic correlations~\cite{Schiffer95}, the specific heat peaks around $T = 1$~K with $C \sim 22$~J/mol K~\cite{Numazawa03}, and hence most of the entropy change will be absorbed to overcome this peak.
Consequently, the lowest achievable temperature using GGG is usually slightly less than 0.5~K.
On the other hand, since GGG still has a quite large specific heat at low temperatures, such as 17~J/mol K at $T = 0.5$~K, the small thermal inflow will not significantly increase the temperature.
Indeed, for the 2~grams of GGG, it is expected that 0.1~K increase around 0.5~K will take roughly one hour with 1~$\mu$W heat inflow.
In reality, there may be several other paths for heat inflow, and the mechanical vibration due to the DC extraction measurement may also increase the temperature, so the heating rate may be faster.

To further reduce the lowest temperature of the present T-ADR system, it may be worthwhile to test several other materials with high magnetic-ion density, listed in the literature~\cite{Wikus14}.
It is also quite interesting to utilize frustrated nature of GGG as suggested earlier~\cite{Zhitomirsky03}.
Future development along these lines may further enhance the capability of the T-ADR system.

\section{Experimental results and discussion}

\begin{figure}
\includegraphics[scale=0.35, angle=-90, trim={0cm 0cm 0cm 0cm}]{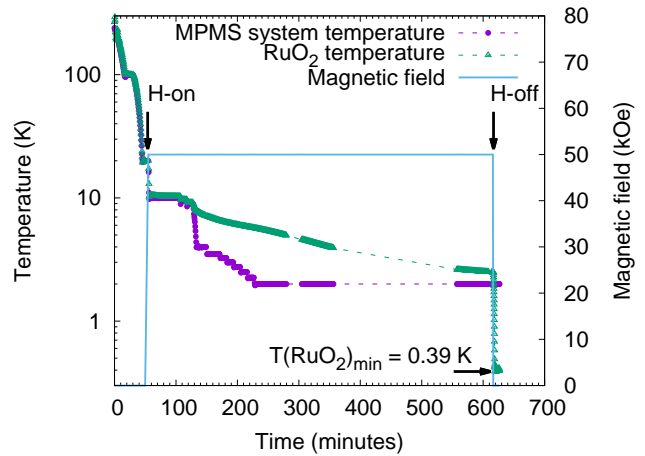}\\
\caption{(Color online) Temperature profile for a typical adiabatic-demagnetization run.
The circles stand for the MPMS system temperature, whereas the triangles for the sample sensor temperature inside the T-ADR system.
Solid line represents the magnetic field; the magnetic field ($H = 50$~kOe) was applied when the MPMS system temperature reached 20~K (approximately 40 minutes from the beginning), whereas it was turned off when the RO sensor temperature became approximately 2.4~K (approximately 10 hours from the beginning).
}
\end{figure}

Figure~3 shows the temperature profile for a typical adiabatic demagnetization run.
At higher temperatures $T > 8$~K, the RO sensor temperature inside T-ADR followed the MPMS system temperature closely.
At 8~K, the RO sensor temperature started to deviate from that of the MPMS system temperature.
This is the sign that the sorption pump (charcoal) starts to work, and the space inside the outer $d \sim 8$~mm straw becomes high vacuum.
The decreasing rate of the RO sensor temperature is typically 0.01~K/min or less, indicating that sufficient thermal insulation is realized with the drinking straw chamber wall and the charcoal sorption pump.
Once the RO sensor temperature reached 2.4~K, the magnetic field was turned off.
Within ten minutes, the RO sensor temperature reached the minimum temperature of $T_{\rm RuO_2} = 0.39$~K.
The standard magnetization measurements using the DC extraction method were then performed with a single extraction repetition.

Since the sample is roughly 95~mm away from the RO sensor, there exists slight temperature gradient.
Hence, we calibrate sample temperature using superconducting transition temperature of polycrystalline zinc.
The procedure is given in Appendix~\ref{appendixTempCalib}.
The deviation of the sample temperature from the RO sensor reading is negligible above 0.8~K, whereas it becomes $\sim 0.08$~K at the lowest temperature.
For the rest of the main text, the sample temperature estimated using the above procedure will be used.

\begin{figure}
\includegraphics[scale=0.35, angle=-90, trim={0cm 0cm 0cm 0cm}]{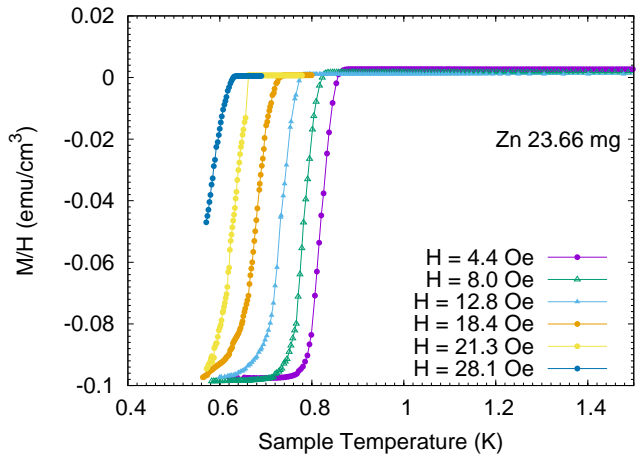}\\
\caption{(Color online) Temperature dependence of $M/H$ of the polycrystalline zinc measured under the external magnetic fields $H = 4.4, 8.0, 12.8, 18.4, 21.3$, and $28.1$~Oe.
Superconducting transition temperatures are estimated from the midpoints.
}
\end{figure}

Figure~4 shows representative results for the magnetization measurements using polycrystalline zinc (99.9999 wt\%, 23.66~mg).
The external magnetic fields were $H = 4.4, 8.0, 12.8, 18.4, 21.3$ and 28.1~Oe.
Procedure for the magnetic-field estimation is given in the Appendix~\ref{appendixTempCalib}.
The superconducting transition of zinc, as well as its suppression under magnetic field, is clearly seen in the figure, indicating typical type-I superconducting behavior.

It should be noted that the superconducting transition in the $H = 21.3$~Oe run is not as smooth as those observed at other fields.
This may be due to the temperature fluctuation; we found typically 0.02~K fluctuation of sample temperature during a single DC extraction scan for the $H = 21.3$~Oe run.
This temperature fluctuations seems to vary from measurement to measurement with the maximum of 0.03~K, plausibly depending on minute difference in sample setting.
Future study for the suppression of this temperature fluctuation is apparently necessary.

Above the superconducting transition temperature, magnetization becomes almost temperature independent.
Slight difference between the data with different external fields for $T > T_{\rm c}$ is due to finite background, which is not a linear function of the external magnetic field.
The fluctuation of the measured magnetization is checked using data obtained at $H = 12.8$~Oe in the temperature range of $1 < T < 1.5$~K; almost all the measured data are in the $\pm 5 \times 10^{-8}$~emu range, with the standard deviation of $\sigma = 2.2 \times 10^{-8}$~emu.
From this result, we can safely say that any magnetic anomaly larger than $1 \times 10^{-7}$~emu may be detected with the present T-ADR system.
This resolution is comparable to that of DC extraction measurements using the standard ($T > 1.8$~K) setup of MPMS.

\begin{figure}
\includegraphics[scale=0.35, angle=-90, trim={0cm 0cm 0cm 0cm}]{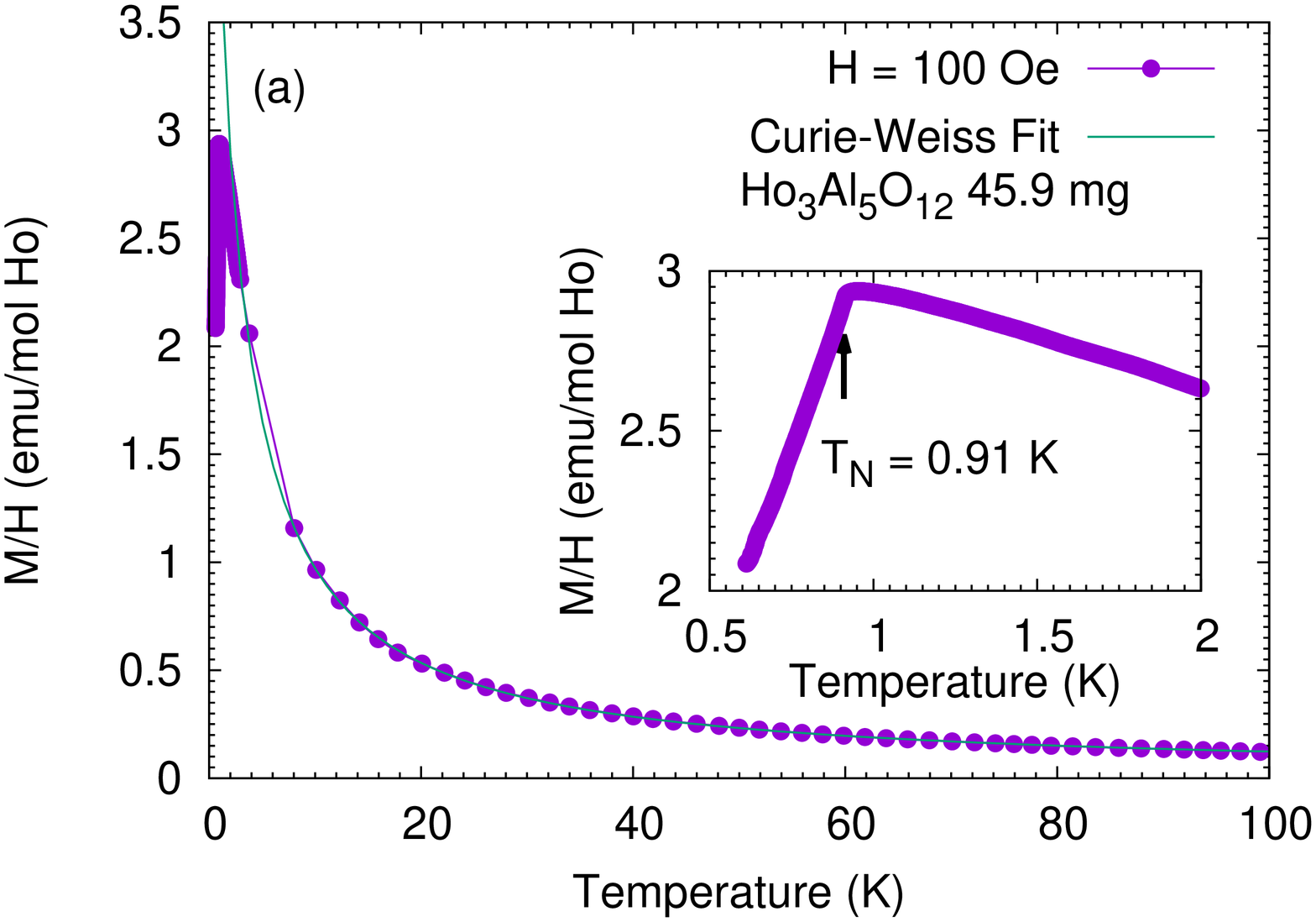}\\
\includegraphics[scale=0.35, angle=-90, trim={0cm 0cm 0cm 0cm}]{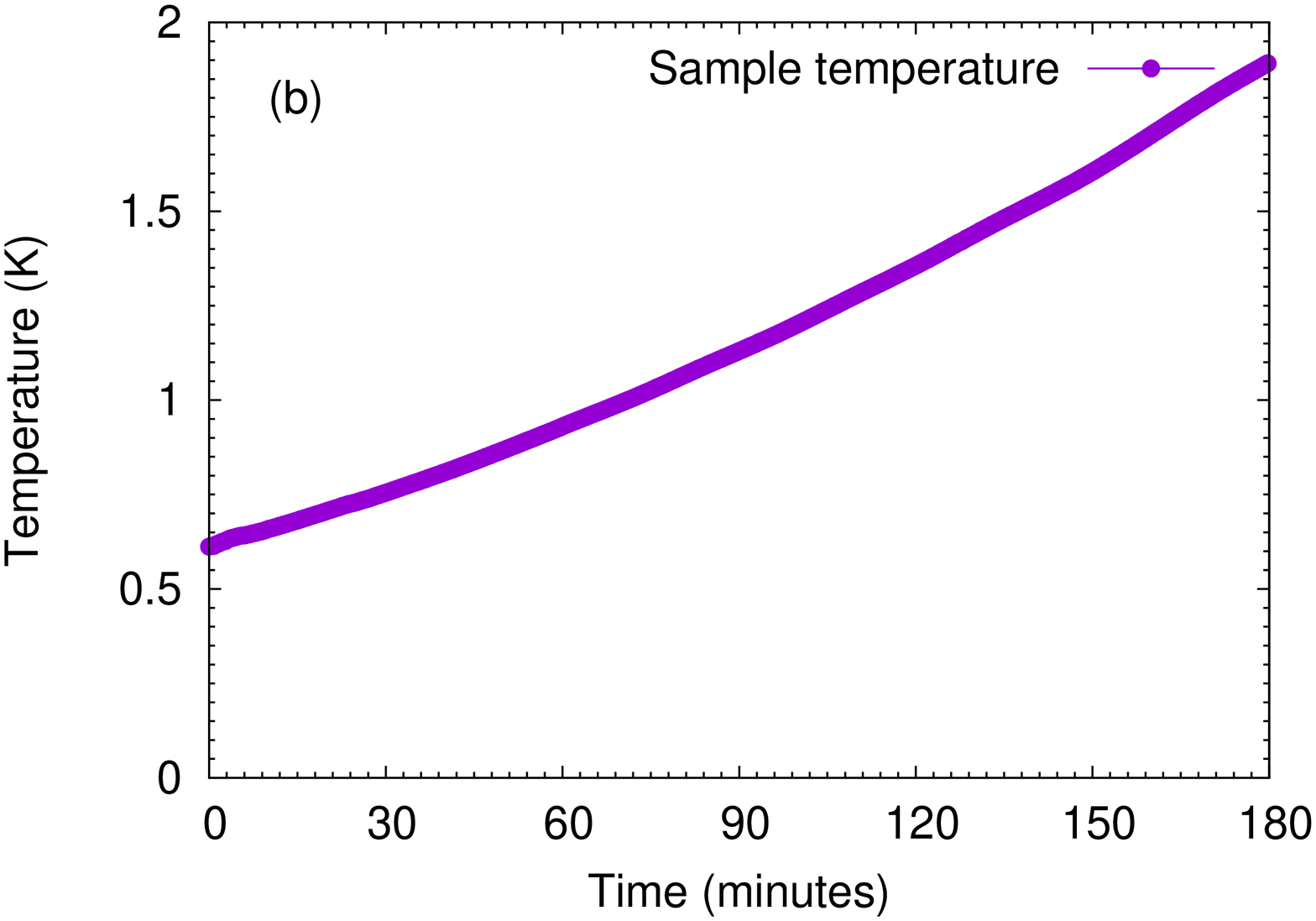}\\
\caption{(Color online) (a) Temperature dependence of $M/H$ of the single crystalline Ho$_3$Al$_5$O$_{12}$ garnet in a wide temperature range up to 100~K.
Solid line is a result of fitting to the Curie-Weiss law.
Details are given in the text.
Inset: magnified figure for the low temperature region $T < 2$~K.
Clear antiferromagnetic transition is seen at $T_{\rm N} = 0.91$~K.
(b) Evolution of the sample temperature in T-ADR during the DC magnetization measurement for Ho$_3$Al$_5$O$_{12}$ as a typical temperature profile.
}
\end{figure}

Figure~5 shows a result of a magnetization measurement on the Ho$_3$Al$_5$O$_{12}$ garnet as a typical antiferromagnet.
The single crystalline sample of 45.9~mg was used in the measurement, and the external field of $H = 100$~Oe was applied.
As shown in Fig.~5(a), typical paramagnetic behavior was observed for $T > 10$~K, which can be perfectly reproduced by the Curie-Weiss law: $\chi = C/(T - T_0) + \chi_0$, where the magnetic susceptibility $\chi$ is defined as $\chi = M/H$, $C$ is the Curie constant, $T_0$ is the Weiss temperature, and $\chi_0$ is the temperature independent term.
The optimal parameters were obtained as: $C = 11.5(1)$~emu/mol K, $T_0 = -2.0(1)$~K, and $\chi_0 = 0.011(1)$~emu/mol.
The effective moment was estimated from $C$ as $\mu_{\rm eff} = 9.60(5)\mu_{\rm B}$, where $\mu_{\rm B}$ stands for the Bohr magneton.
This estimated effective moment is in semi-quantitative agreement with that of the free Ho$^{3+}$ ions ($10.6\mu_{\rm B}$); the slight discrepancy may be due to the crystalline electric field splitting~\cite{Nagata01}.
The data in the lower temperature range are shown in the inset of Fig.~5(a).
The clear anomaly was observed at $T_{\rm N} = 0.91$~K.
The transition temperature in the earlier reports fluctuates slightly; 0.95~K in the neutron diffraction experiment~\cite{Hammann69}, 0.85~K in the magnetic susceptibility measurement~\cite{Cooke67}, and 0.839~K in the specific heat measurement~\cite{Nagata01}.
Our $T_{\rm N}$ is in the fluctuation range.
Temperature profile during the magnetization measurement is shown in Fig.~5(b).
The temperature increase is quite slow as $\sim 0.008$~K/min.

\begin{figure}
\includegraphics[scale=0.35, angle=-90, trim={0cm 0cm 0cm 0cm}]{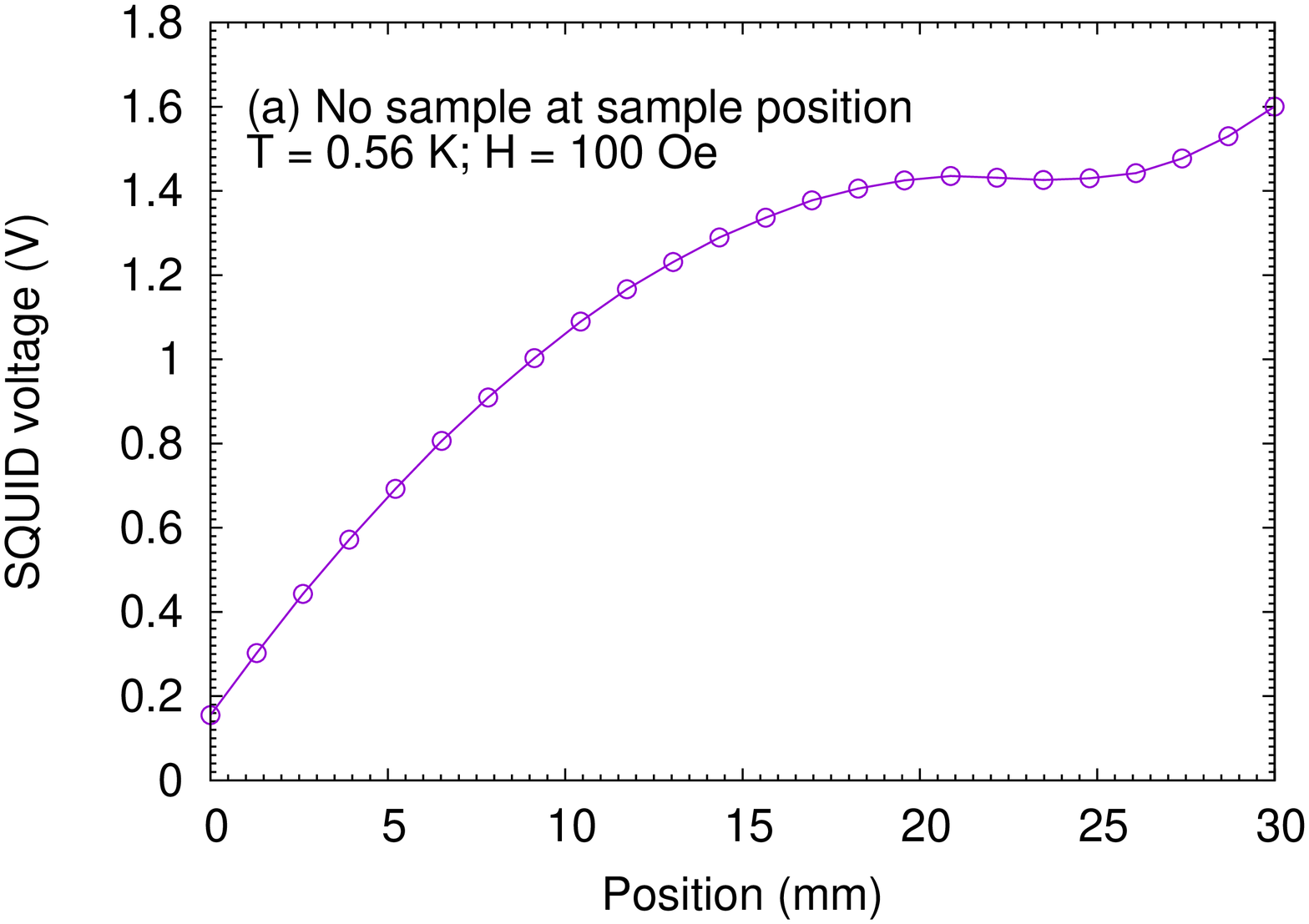}\\
\includegraphics[scale=0.35, angle=-90, trim={0cm 0cm 0cm 0cm}]{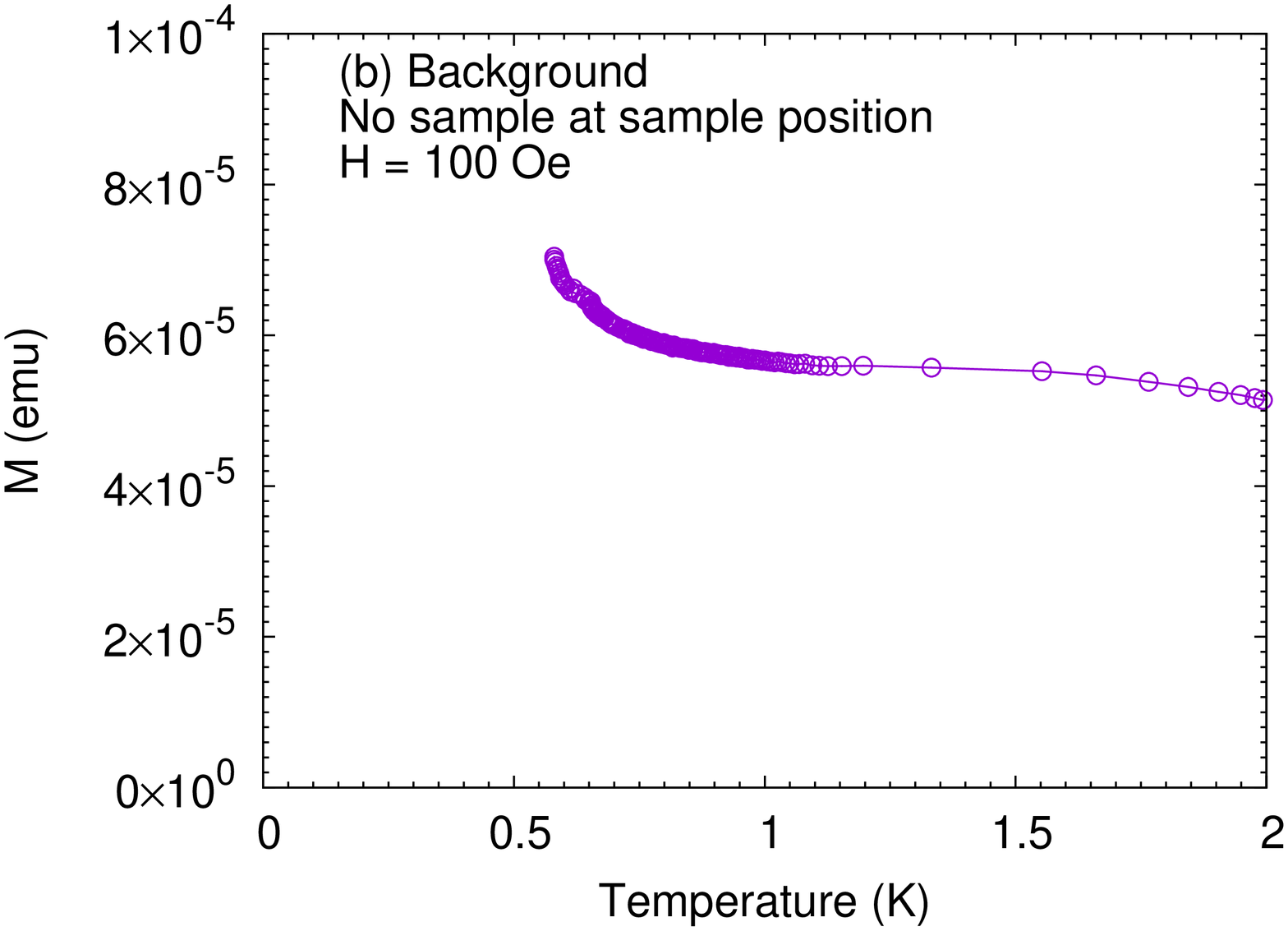}\\
\caption{(Color online) (a) SQUID output for the DC extraction scan with no sample at the sample position.
The external field of $H = 100$~Oe is applied.
In addition to the linear asymmetric signal, there exists a broad peak centered around the 15~mm position, which results in fictitious (background) signal in magnetization measurements.
(b) Temperature dependence of the fictitious background signal measured under $H = 100$~Oe with no sample at the sample position.
}
\end{figure}

In this T-ADR setup, the magnetic refrigerant is situated at 110~mm above the sample position, and hence the refrigerant is out of the MPMS superconducting magnet when sample magnetization is measured.
Nonetheless, stray magnetic field from the MPMS magnet can magnetize the refrigerant, and the magnetic flux variation at the SQUID detection coil due to the motion of the refrigerant gives rise to significant fictitious signal in the DC extraction measurements.
Typical SQUID signal as the DC transport was moved for 30~mm is shown in Fig.~6(a); no sample was set at the sample position for this scan.
(At 15~mm, the sample position of T-ADR coincides with the center of the MPMS magnet.)
An asymmetric SQUID output with a weak peak around $\sim 15$~mm can be seen in the figure, which, by using a standard analysis procedure for DC extraction measurements, is recognized as finite background signal.
Figure~6(b) shows the temperature dependence of thus-obtained background signal.
The center position for the extraction scan was fixed to the sample position.
The background is reasonably small as $\sim 5 \times 10^{-5}$~emu at $H = 100$~Oe.
There appears slight increase of the background as the temperature is decreased, which may be attributed to the increase of the paramagnetic susceptibility of the magnetic refrigerant.
In this report, to show the bare performance of the magnetization measurement using the T-ADR system, we did not subtract the background from the raw data.
However, because the background is only weakly temperature dependent except for the lowest temperatures, the subtraction of the background is straightforward.

\section{Conclusions}

In the present study, we have successfully developed a tiny adiabatic-demagnetization refrigerator (T-ADR) to be used in combination with a commercial SQUID-based magnetometer.
It was confirmed that low temperature magnetization can be successfully measured using T-ADR down to $\sim 0.5$~K.
The fluctuation of the measured magnetization is within $\pm 5 \times 10^{-8}$~emu, which is almost the same as that of the standard DC extraction method of MPMS magnetometer.
On the other hand, relatively large background of $5 \times 10^{-5}$~emu was observed, which is largely due to the paramagnetic response of the magnetic refrigerant.
The diamagnetic signal of the superconducting zinc was successfully measured up to $H = 21.5$~Oe, whereas the antiferromagnetic transition in Ho$_3$Al$_5$O$_{12}$ was clearly observed under $H = 100$~Oe.
Those results reasonably reproduce the earlier studies, confirming the reliability of the magnetization data obtained with the present T-ADR setup.
It is quite straightforward to implement electrical resistivity and Hall resistivity measurements at ultra-low temperatures using the present T-ADR system.

\begin{acknowledgments}
The authors thank T. Onimaru for suggesting the possibility of low-temperature magnetic measurements using the ADR technique.
This work was partly supported by Grants-in-Aid for Scientific Research (24224009 and 23244068) from MEXT of Japan, and by the Nano-Macro Materials, Devices and System Research Alliance.
\end{acknowledgments}

\appendix
\section{Estimation of temperature increase under finite magnetic field}\label{appendixTempField}
\begin{figure}
\includegraphics[scale=0.35, angle=-90, trim={0cm 0cm 0cm 0cm}]{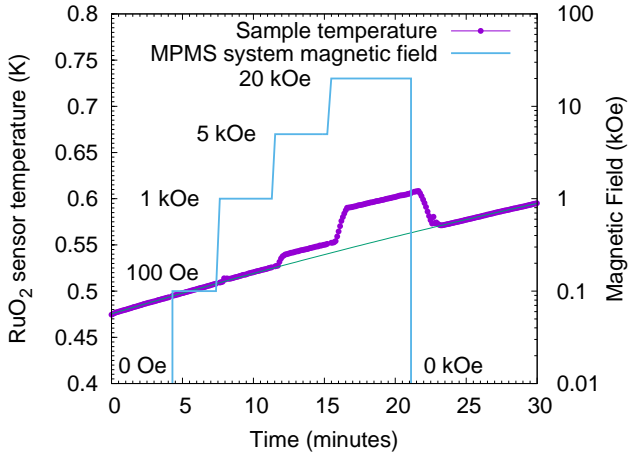}\\
\caption{(Color online) Evolution of the RO sensor temperature as the external magnetic field was increased.
After the sample was set to the measuring position ({\it i.e.}, center of the MPMS superconducting magnet), the RO sensor temperature was recorded with the stepwise increase of the magnetic field with the steps being $H = 0.1, 1, 2, 5$ and 20~kOe.
Green solid line stands for an estimate of temperature evolution under zero external magnetic field, whereas blue solid line for the MPMS system magnetic field.
It may be noted that because of the finite sweeping rate of the superconducting magnet, there is a time lag for the actual magnetic field to be set to the MPMS system magnetic field.
This is the reason for the apparent delay in the behavior of the RO sensor temperature from that of the MPMS system magnetic field.
}
\end{figure}
The temperature increase under finite magnetic field was checked with the magnetic refrigerant set at the measuring position.
Figure~7 shows the evolution of the RO sensor temperature under external magnetic fields of $H = 0.1, 1, 5$ and 20~kOe, measured as the temperature gradually increases.
It may be noted that up to $H = 1$~kOe, the temperature increase is hardly seen; even for $H = 5$~kOe, only a slight change of $\Delta T_{\rm RuO_2} = 0.01$~K was seen.
At $H = 20$~kOe, a sizable increase of $\Delta T_{\rm RuO_2} = 0.06$~K was observed.
Hence, we can safely say that the lowest temperature of T-ADR is not significantly affected by the external magnetic field up to 5~kOe.

\section{Sample position temperature}\label{appendixTempCalib}
The sample position temperature was estimated using the superconducting transition temperature of the polycrystalline zinc.
The raw (temperature-uncalibrated) data are given in Fig.~8, which were obtained under MPMS-system magnetic fields set to $H_{\rm set} = 3, 6, 10, 15, 20$ and 25~Oe.
Real magnetic fields applied to the sample were estimated as follows.
External field was directly measured using the fluxgate for the $H_{\rm set} = 6.0$~Oe as $H = 8.0$~Oe, whereas those for $H_{\rm set} = 20$ and 25~Oe were estimated as $H = 21.3$ and 28.1~Oe using the Pd standard paramagnetic signal.
Fields for $H_{\rm set} = 3, 10$ and 15~Oe were estimated as $H = 4.4, 12.8$ and $18.4$~Oe by normalizing the perfect diamagnetic signal of zinc to that obtained at $H = 8.0$~Oe.
The superconducting transition temperatures in the uncalibrated temperature scale were estimated from the midpoints, and were used to estimate the real sample position temperatures using the well-established $H_{\rm c}(T)$ function of zinc~\cite{Seidel58}.
It should be noted that the transition temperatures in the raw (uncalibrated) temperature scale at lower fields, such as $H = 4.4$~Oe ($T_{\rm c}^{\rm raw} = 0.812$~K) and 8.0~Oe ($T_{\rm c}^{\rm raw} = 0.769$~K), well extrapolate to the tabulated value $T_{\rm c} = 0.851$~K for $H \rightarrow 0$~\cite{EPT76}.
This indicates that the temperature gradient between the RO sensor position and the sample position is negligible at $T \sim 0.8$~K.
Estimated sample temperatures, using zinc superconducting transition, were plotted as a function of the RO sensor temperature in the inset of Fig.~8.
Roughly 0.08~K difference can be seen between the sample temperature and RO sensor temperature at $T_{\rm RuO_2} \sim 0.48$~K.
Using a polynomial fitting, a correction function was made to convert the RO sensor temperature to the sample temperature, and was used to obtain sample temperature for all the data shown in Figs. 4 to 6.

\begin{figure}
\includegraphics[scale=0.35, angle=-90, trim={0cm 0cm 0cm 0cm}]{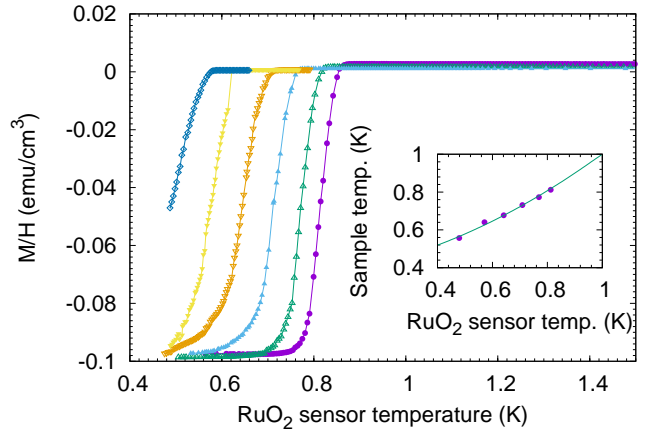}\\
\caption{(Color online) Measured magnetization of the polycrystalline zinc given in Fig.~4 as a function of original RO sensor temperature. 
The same legends as in Fig.~4 are used.
Inset: RO sensor temperature versus actual sample temperature.  
The sample temperature is estimated from the superconducting transition temperature of zinc under each given external magnetic field.
}
\end{figure}

%

\end{document}